\documentclass[mathleft,fleqn,%
]{an}
\usepackage{graphicx}
\usepackage[varg]{txfonts}
\begin{document}

\Pagespan{1}{}
\Yearpublication{2017}%
\Yearsubmission{2016}%
\Month{0}%
\Volume{999}%
\Issue{0}%
\DOI{asna.201400000}%


\title{Wide $\sigma$~Orionis binaries resolved by UKIDSS}

\author{Jos\'e A. Caballero\inst{1,2}\fnmsep\thanks{\tt caballero@cab.inta-csic.es}
\and I. Novalbos\inst{3}
\and T. Tobal\inst{3}
\and F. X. Miret\inst{3}
}
\titlerunning{Wide $\sigma$~Orionis binary stars resolved by UKIDSS}
\authorrunning{J.\,A. Caballero et~al.}
\institute{
Centro de Astrobiolog\'{\i}a (INTA-CSIC), European Space Astronomy Centre campus, Camino Bajo del Castillo, E-28691 Villanueva de la Ca\~nada, Madrid, Spain
\and Landessternwarte K\"onigstuhl, Zentrum f\"ur Astronomie der Universit\"at Heidelberg, K\"onigstuhl 17, D-69117 Heidelberg, Germany
\and Observatori Astron\`omic del Garraf, Barcelona, Spain
}

\received{2016 Nov 27}
\accepted{2017 Jun 26}
\publonline{2017 Mmm dd}


\abstract{%
In spite of its importance for the study of star formation at all mass domains, the nearby young $\sigma$~Orionis cluster still lacks a comprehensive survey for multiplicity.
We try to fill that observational gap by looking for wide resolved binaries with angular separations between 0.4 and 4.0\,arcsec.
We search for companions to {331} catalogued cluster stellar members and candidates in public $K$-band UKIDSS images outside the innermost 1\,arcmin, which is affected by the glare of the bright, eponymous $\sigma$~Ori multiple system, and investigate their cluster membership with colour-magnitude diagrams and previous knowledge of youth features.
Of the 18 identified pairs, ten have very low {individual} probabilities of chance alignment ($<$~1\,\%) and are considered here as physical pairs.
Four of them are new, while the other six had been discovered previously, but never investigated homogeneously and in detail.
Projected physical separations and magnitude differences of the ten probably bound pairs range from 180 to 1220\,au, and from 0.0 to 3.4\,mag in $K$, respectively.
Besides, we identify two cluster stars with elongated point spread functions.
We determine the minimum frequency of wide multiplicity in the interval of projected physical separations $s$ = 160--1600\,au in $\sigma$~Orionis at {3.0$^{+1.2}_{-1.1}$\,\%}.
We {discover} a new Lindroos system, find that massive and X-ray stars {tend} to be in pairs or trios, conclude that multiplicity truncates circumstellar discs {and} enhances X-ray emission, and ascribe a reported lithium depletion in a young star to unresolved binarity in spectra of moderate resolution. 
When accounting for all know multiples, including spectroscopic binaries, the minimum frequency of multiplicity increases to about 10\,\%, which implies that of the order of 80--100 unknown multiple systems still await discovery in $\sigma$~Orionis.
}%

\maketitle


\section{Introduction}
\label{section.introduction}

Stars can be single or members in double and multiple systems.
Multiplicity is, together with mass function, disc frequency, and spatial distribution, one of the key parameters for constraining low-mass star-formation scenarios (Reipurth \& Clarke 2001; Chabrier 2003; Lada 2006; Goodwin et~al. 2007; Bate 2009; Parker et~al. 2009).
There have been numerous surveys for multiplicity in the field (e.g., Duquennoy \& Mayor 1991; Reid \& Gizis 1997; Bouy et~al. 2003; Raghavan et~al. 2010; Janson et~al. 2012) and in open clusters (see the review by Duch\^ene \& Kraus 2013 and references therein).
Virtually every relevant open cluster younger than $\sim$600\,Ma and at less than 1\,kpc has been surveyed for multiplicity, both resolved (i.e., by means of speckle, first, and adaptive optics and space missions, next) and unresolved (i.e., spectroscopic binarity):
Hyades (Mason et~al. 1993a; Patience et~al. 1998),
$\alpha$~Persei and Praesepe (Mason et~al. 1993b; Patience et~al. 2002),
Pleiades (Steele \& Jameson 1995; Bouvier et~al. 1997; Mart\'{\i}n et~al. 2000; {Bouy et~al. 2006a}; Lodieu et~al. 2007),
the Scorpius-Centaurus-Lupus-Crux complex (including Upper Scorpius -- Kraus et~al. 2005; {Bouy et~al. 2006b}; Kouwenhoven et~al. 2007; {Lafreni\`ere et~al. 2014}),
Taurus-Auriga (Ghez et~al. 1993; Leinert et~al. 1993; Kraus et~al. 2011),
the Orion Nebula Cluster (Padgett et~al. 1997; Preibisch et~al. 1999; K\"ohler et~al. 2006; Reipurth et~al. 2007), 
$\rho$~Ophiuchi (Ratzka et~al. 2005), 
or Chamaeleon~I (Lafreni\`ere et~al.~2008). 

The \object{$\sigma$~Orionis} open cluster ($d =$ 387.5$\pm$1.3\,pc, $\tau \sim$ 3\,Ma) in the Ori~OB1b association is, especially because of its numerous and well-investigated brown-dwarf and planetary-mass population, another cornerstone cluster for the study of the stellar and substellar formation (Garrison 1967; Wolk 1996; B\'ejar et~al. 1999; Zapatero Osorio et~al. 2000; Caballero 2008c; Schaefer et~al. 2016). 
Therefore, it should have its own comprehensive multiplicity survey. 
However, this study is still missing.
{Caballero (2014) tried to fill part of this observing gap with a literature review on multiplicity in $\sigma$~Orionis, but a reliable, exhaustive, homogeneous survey does not exist yet.}

In this work, we focus on wide binaries in $\sigma$~Orionis that can be resolved from the ground with standard imaging (i.e., no lucky or adaptive-optics imaging), but that may have been uncovered by previous all-sky surveys such as DENIS or 2MASS (Epchtein et~al. 1997; Skrutskie et~al. 2006).
This requisite translates into exploring angular separations between 0.4 and 4.0\,arcsec approximately, or, alternatively, projected physical separations between 160 and 1600\,au at the most probable cluster distance.
To achieve our goal, we used data of the UKIRT Infrared Deep Sky Survey (UKIDSS; Lawrence et~al. 2007).

\section{Analysis and results}
\label{section.analysis+results}

   \begin{table*}
      \caption[]{New (upper part) and discarded (lower part) Mayrit stars$^a$.} 
         \label{table.mayrit}
     $$ 
         \begin{tabular}{ll ccc l}
            \hline
            \hline
            \noalign{\smallskip}
Mayrit  			& Alternative 					& $\alpha$ 	& $\delta$      	& $J_{\rm 2M}$		& References$^{b}$		 	\\
	  			& name						& (J2000)		& (J2000)       	& [mag]			& 						\\
            \noalign{\smallskip}
            \hline
            \noalign{\smallskip}
{68229}			& [W96] pJ053941--0236			& 05 38 41.36   & --02 36 44.5  & 12.30$\pm$0.02	& Wo96, Ca10b  			\\ 
{168291} AB		& [W96] rJ053834--0234			& 05 38 34.31   & --02 35 00.1  & 11.22$\pm$0.03	& Wo96, Ca10b, He14  		\\ 
{172264}			& [SWW2004] 130 				& 05 38 33.36   & --02 36 17.6  & 12.05$\pm$0.03	& Sh04, Ca10b  			\\ 
{270196}			& [HHM2007] 655			 	& 05 38 39.73   & --02 40 19.7  & 13.75$\pm$0.03	& He07, Ca10b  			\\ 
{332167}			& [SWW2004] 200				& 05 38 49.93 	& --02 41 22.8	& 12.75$\pm$0.03	& Sh04, Sa08, Ca10b, He14  	\\ 
{492211}			& [W96] rJ053827--0242			& 05 38 27.74   & --02 43 01.0  & 12.19$\pm$0.03	& Wo96, Sa08, Ca10b, He14  	\\ 
{605079}			& [SWW2004] 127 				& 05 39 24.36   & --02 34 01.4  & 12.98$\pm$0.03	& Sh04, Sa08, Ca10b  		\\ 
{957055} 			& [SWW2004] 163			 	& 05 39 37.30   & --02 26 56.8  & 11.70$\pm$0.03	& Sh04, Ca10b, Ca12, He14  	\\ 
{1093033}		& [HHM2007] 1030			 	& 05 39 24.56   & --02 20 44.1  & 11.37$\pm$0.03	& He07, Ca10b, He14  		\\ 
{1149270}		& [HHM2007] 59 				& 05 37 28.07	& --02 36 06.6  & 13.74$\pm$0.03	& He07, LSC08$^{c}$		\\ 
{1160240}		& [SWW2004] 184 				& 05 37 37.85	& --02 45 44.2  & 12.69$\pm$0.03	& Sh04, LSC08, He14	 	\\ 
{1178039}		& [SWW2004] 138 				& 05 39 33.79   & --02 20 39.9  & 12.37$\pm$0.03	& Sh04, Ca10b  			\\ 
{1344302}		& [SWW2004] 172				& 05 37 28.32	& --02 24 18.2  & 14.00$\pm$0.03	& Sh04, LSC08	 			\\ 
            \noalign{\smallskip}
            \hline
            \noalign{\smallskip}
(111208)			& UCM 0536--0239				& 05 38 41.24	& --02 37 37.7  & 16.38$\pm$0.09	& Ca07b, Ca08b			\\ 
(258215)			& [W96] pJ053834--0239			& 05 38 34.79	& --02 39 30.0  & 10.44$\pm$0.03	& Wo96, Sa08, He14		\\ 
{(459340)}		& StHA 50						& 05 38 34.44	& --02 28 47.6	& 10.67$\pm$0.02	& {Ste86, Ca08b, Ca17}		\\ 
(552137)			& [HHM2007] 918				& 05 39 10.04	& --02 42 42.5  & 12.97$\pm$0.03	& He07, Sa08, He14	 		\\ 
(660067)			& [W96] pJ053925--0231			& 05 39 25.34	& --02 31 43.7  & 12.43$\pm$0.03	& Wo96, Sa08, He14		\\ 
{(717307)}		& TYC 4771--950--1				& 05 38 06.49	& --02 28 49.4  & 10.09$\pm$0.03	& {Fr06, Ca17}				\\ 
(926051)			& UCM 0537--0227				& 05 39 32.70	& --02 26 15.4  & 16.14$\pm$0.10	& Ca08b, Lo09				\\ 
{(1227243)}		& HD 294275					& 05 37 31.87	& --02 45 18.4  &   9.26$\pm$0.03	& {Ca17}					\\ 
{(1456284)}		& TYC 4770--1261--1			& 05 37 10.49	& --02 30 07.0  & 10.72$\pm$0.03	& {LSC08, Ca17}			\\ 
{(1468100)}		& HD 294301					& 05 40 21.12	& --02 40 25.5  & 10.21$\pm$0.03	& {GC93, Ca17}			\\ 
(1659068)		& HD 294297					& 05 40 27.54	& --02 25 43.1  &   9.06$\pm$0.03	& GH08, Ca10a			\\ 
         \noalign{\smallskip}
            \hline
         \end{tabular}
     $$ 
\begin{list}{}{}
\item[$^{a}$] {Objects in the Mayrit catalogue of stars and brown dwarfs in the $\sigma$~Orionis cluster outside the innermost 1\,arcmin that were not listed by Caballero (2008c) and with both DENIS $i$ and 2MASS $JHK_s$ detections [new, upper part], or that were in the original catalogue but have been classified as non-cluster members afterwards [discarded, bottom part].
Mayrit numbers of discarded objects (in parenthesis), either stars or galaxies, shall not be used.}
\item[$^{b}$] {References -- 
{Ste86: Stephenson 1986;
GC93: Gray \& Corbally 1993;}
Wo96: Wolk 1996;
Sh04: Sherry et~al. 2004;
{Fr06: Franciosini et~al. 2006;}
Ca07b: Caballero 2007b; 
He07: Hern\'andez et~al. 2007;
Ca08: Caballero et~al. 2008b;
GH08: Gonz\'alez Hern\'andez et~al. 2008;
LSC08; L\'opez-Santiago \& Caballero 2008;
Ma08: Maxted et~al. 2008;
Sa08: Sacco et~al. 2008;
Lo09: Lodieu et~al. 2009;
Ca10a: Caballero 2010a; 
Ca10b: Caballero et~al. 2010b; 
Ca12: Caballero et~al. 2012;
Ca14: Caballero 2014;
He14: Hern\'andez et~al. 2014};
{Ca17: Caballero 2017.}
\item[$^{c}$] {\object{Mayrit~1149270} was classified as a cluster non-member by Hern\'andez et~al. (2014).
However, it is a strong X-ray emitter in the cluster with slightly blue colours according to L\'opez-Santiago \& Caballero (2008).} 
\end{list}
   \end{table*}

We downloaded public UKIDSS $K$-band images from the WFCAM Science Archive of {331} $\sigma$~Orionis cluster members and member candidates tabulated in the Mayrit catalogue (Caballero 2008c) and subsequent additions {and corrections (see below)}.
In particular, we got square images, 300\,pixel in side, from the UKIDSS Eighth Data Release DR8+ taken under the Galactic Clusters Survey (GCS; Lodieu et~al. 2009).
In spite of UKIRT WFCAM (the camera with which the UKIDSS survey was carried out; Casali et~al. 2007) having a pixel scale of 0.400$\pm$0.010\,arcsec, the $K$-band images had a pixel scale of only 0.200$\pm$0.005\,arcsec. 
This improved sampling of the median seeing at UKIRT ($\sim$\,0.7\,arcsec) was thanks to the interlacing of data obtained in a 2$\times$2 microstepping sequence with step size $N$+0.5\,pixels (cf. Lawrence et~al. 2007).
As a result, the angular size of the downloaded $K$-band images was of 60\,arcsec.
Since the microstepping method was applied to the $K$ GCS images only, the ones in $ZYJH$ bands had the original pixel scale of 0.4\,arcsec.
All UKIDSS observations in the $\sigma$~Orionis area were performed either on 2005 Oct 14 (epoch 2005.79) or 2005 Nov 22 (epoch 2005.89).
Gaussian full-width half maxima of most non-saturating single stars within 30\,arcmin to the $\sigma$~Ori multiple system ranged between 0.6 and 1.0\,arcsec in the $K$ band.

Although the membership in cluster of many stars and brown dwarfs in the {almost-ten-year-old} Mayrit catalogue has been confirmed or refuted in newer {astrometric} (Caballero 2010a, {2017}), X-rays (L\'opez-Santiago \& Caballero 2008; Caballero et~al. 2010b), and spectroscopic surveys (Caballero et~al. 2008b, 2012; Maxted et~al. 2008; Sacco et~al. 2008; Hern\'andez et~al. 2014; Koenig et~al. 2015), it is still the most comprehensive catalogue of cluster members and member candidates in $\sigma$~Orionis.
The Mayrit catalogue is complete down to (and slightly beyond) the cluster substellar boundary at $J \sim$ 14.5\,mag (Caballero et~al. 2007), where the UKIDSS quality image is still good enough for multiplicity studies.
An exhaustive analysis of multiplicity of much fainter brown dwarfs and planetary-mass objects in $\sigma$~Orionis would require a new deeper dataset, such as {VIRCAM (Pe\~na Ram\'{\i}rez et~al. 2012; V.~J.~S. B\'ejar, in prep.) or VIMOS (Elliott et~al. 2017).}
Except for the innermost arc\-minute in the cluster centre, which is affected by the intense glare of the bright eponymous $\sigma$~Ori trapezium, both the UKIDSS and Mayrit catalogues nicely complement each other for wide multiplicity studies. 
For completeness, we updated the Mayrit catalogue with 13 new additions and {11} deletions.
The names and coordinates of the {24} new/discarded stars and galaxies, together with references for discovery, non-/membership feature identification, and Mayrit number\footnote{The Mayrit number provides the separation to the cluster centre in arcsec and position angle in deg.
For example, Mayrit~92149 is at $\rho$ = 92\,arcsec and $\theta$ = 149\,deg to $\sigma$~Ori\,Aa,Ab,B,  Mayrit~359179 is at $\rho$ = 359\,arcsec and $\theta$ = 179\,deg, etc.}, are provided in Table~\ref{table.mayrit}.
The final star counting is as follows: 
338 sources in the original Mayrit catalogue, --9 stars inside the central arcminute, +13 additions, --{11} deletions, which make {331} investigated {cluster} stars.

Once we downloaded all the UKIDSS $K$-band images, we looked {by eye} for resolved companions at $\rho <$ 4\,arcsec to the {331} Mayrit targets outside the central arc\-minute.
{For completeness, we did the same for the 11 deletions.}
We choose $\rho$ = 4\,arcsec as the search boundary because the best 2MASS images had full-width-half-maxima of 2.5\,arcsec and the 2MASS Point Source Catalog ``standard aperture'' magnitude was measured in a 4\,arcsec-radius aperture (cf. Skrutskie et~al. 2006).
{Wider companions brighter than the 2MASS completeness magnitudes should have been detected in previous comprehensive surveys in the area}.
In Table~\ref{table.physical+optical} and Fig.~\ref{figure.estrellitas} we show the 18 identified multiple system candidates.
We were interested on sources with stellar point spread functions (PSFs) only and, therefore, we did not account for a faint background extended optical companion, probably of extragalactic nature, at 2--3\,arcsec to the NWN to Mayrit~270196.
As expected, none of the 18 pairs had been resolved by 2MASS.
{Four of them had 2MASS $J$- and $H$-band detections where the goodness-of-fit quality of the profile-fit photometry was very poor, which is indicated by the category `EEA' in the photometric quality flag.
One star, namely TYC~4770--1261--1, had a very poor fit also in the $K_s$ band ({\tt Qflg} = `EEE').}

   \begin{table*}
      \caption[]{Names, astrometry, and photometry of candidate physical (upper part) and optical (lower part) binaries resolved in the UKIDSS images outside the innermost 1\,arcmin.} 
         \label{table.physical+optical}
     $$ 
         \begin{tabular}{ll ccc cccc}
            \hline
            \hline
            \noalign{\smallskip}
Mayrit  			& Alternative 				& $\alpha_{\rm 2M}$ & $\delta_{\rm 2M}$ & $J_{\rm 2M}$		& $\rho$			& $\theta$      		& $\Delta K_{\rm UKIDSS}$ & Prob.$^a$	\\
	  			& name					& (J2000)			& (J2000)       		& [mag]			& [arcsec]			& [deg]       		& [mag]				& [\%]		\\
            \noalign{\smallskip}
            \hline
            \noalign{\smallskip}
{92149} AB$^{b,c}$	& [W96] rJ053847--0237 		& 05 38 47.92   & --02 37 19.2  & 12.02$\pm$0.04			& 2.14$\pm$0.02	&  63.7$\pm$0.5	& 0.8274$\pm$0.0012	& 0.5		\\ %
{359179} AB		& V595 Ori (Haro 5--14)  		& 05 38 45.38   & --02 41 59.4  & 11.99$\pm$0.03			& 0.97$\pm$0.08  	& 225$\pm$4		& 0.90$\pm$0.03		& 0.02	\\ %
{489165} AB		& [W96] rJ053853--0243  		& 05 38 53.17   & --02 43 52.8  & 12.24$\pm$0.03			& 1.26$\pm$0.02   	& 356.8$\pm$0.9	& 1.824$\pm$0.003		& 0.03	\\ %
{707162} AB$^{b}$	& [W96] rJ053859--0247  		& 05 38 59.11   & --02 47 13.3  & 11.32$\pm$0.03			& 0.73$\pm$0.02    	& 89.3$\pm$1.6  	& 0.6808$\pm$0.0012	& 0.007	\\ %
{785038} AB		& Kiso A--0904 80			& 05 39 17.18   & --02 25 43.4  & 12.90$\pm$0.03			& 0.48$\pm$0.08  	& 333$\pm$4		& 0.02$\pm$0.04		& 0.003	\\ %
{960106} AB		& V1147 Ori (HD 37633)  		& 05 39 46.20   & --02 40 32.1  &  8.88$\pm$0.04			& 3.17$\pm$0.02   	& 232.3$\pm$0.4	& 3.41$\pm$0.02		& 0.09	\\ %
{1106058} AB$^b$& 2E 1486					& 05 39 47.42   & --02 26 16.3  & 10.04$\pm$0.03			& 2.75$\pm$0.02   	& 219.9$\pm$0.4	& 1.9887$\pm$0.0011	& 0.06	\\ %
{1245057} AB$^{b,c}$& V605 Ori (Haro 5--31) 		& 05 39 54.29   & --02 24 38.6  & 12.34$\pm$0.03			& 2.16$\pm$0.02   	& 184.2$\pm$0.5	& 0.0250$\pm$0.0008	& 0.03	\\ %
{1411131} AB$^{b,c}$& [SWW2004] 118	 		& 05 39 56.02   & --02 51 22.8  & 12.04$\pm$0.04			& 1.98$\pm$0.02   	& 300.6$\pm$0.6	& 0.9759$\pm$0.0015	& 0.03	\\ 
{1564349} AB$^b$	& 2E 1464				& 05 38 25.50   & --02 10 22.9  & 12.11$\pm$0.03			& 0.86$\pm$0.04   	& 48$\pm$2		& 0.24$\pm$0.03		& 0.004	\\ %
            \noalign{\smallskip}
            \hline
            \noalign{\smallskip}
{168291} AB$^{b,c}$& [W96] rJ053834--0234 		& 05 38 34.31	& --02 35 00.1  & 11.22$\pm$0.03			& 3.28$\pm$0.02	&  55.2$\pm$0.4	& 2.924$\pm$0.003		& ...		\\ %
{433123} AB		& V2740 Ori (S\,Ori 25)  		& 05 39 08.95   & --02 39 58.0  & 14.66$\pm$0.03			& 3.78$\pm$0.02	& 38.3$\pm$0.3	& 2.63$\pm$0.04		& ...		\\ 
{803197} AB		& S\,Ori J053829.0--024847	& 05 38 28.97 & --02 48 47.3  & 14.82$\pm$0.04			& 1.49$\pm$0.02   	& 249.3$\pm$0.8	& 3.75$\pm$0.12		& ...		\\ %
{1207010} AB		& [SWW2004] 196	 		& 05 38 58.32   & --02 16 10.1  & 12.34$\pm$0.03			& 2.74$\pm$0.02   	&  55.8$\pm$0.4	& 4.122$\pm$0.014		& ...		\\ %
{(1456284)} AB$^{c,d}$& TYC 4770--1261--1		& 05 37 10.47   & --02 30 07.2  & 10.72$\pm$0.05			& 3.49$\pm$0.02   	&  59.5$\pm$0.3	& 1.0014$\pm$0.0010	& ...		\\ 
{1610344} AB$^b$	& [SE2004] 6		 		& 05 38 14.54	& --02 10 15.3  & 12.48$\pm$0.03			& 3.82$\pm$0.02   	& 141.4$\pm$0.3	& 4.82$\pm$0.04		& ...		\\ 
{1691180} AB		& [SWW2004] 64	 		& 05 38 43.80   & --03 04 11.5  & 12.94$\pm$0.03			& 3.74$\pm$0.02   	& 342.9$\pm$0.3	& 3.496$\pm$0.017		& ...		\\ %
{1788137} AB		& [SWW2004] 58	 		& 05 40 06.76   & --02 57 38.9  & 13.65$\pm$0.04			& 1.39$\pm$0.02   	& 265.1$\pm$0.8	& 0.529$\pm$0.004		& ...		\\ %
         \noalign{\smallskip}
            \hline
         \end{tabular}
     $$ 
\begin{list}{}{}
\item[$^{a}$] {Chance alignment probabilities. See Section~\ref{section.discussion}.}
\item[$^{b}$] {Previously known pairs. 
See details in Caballero (2014).}
\item[$^{c}$] {2MASS quality flag EEA or EEE.}
\item[$^{d}$] {Discarded cluster member candidate in Table~\ref{table.mayrit}.}
\end{list}
   \end{table*}

For the 18 pairs, we report accurate relative astrometry ({angular separation} $\rho$ and {position angle} $\theta$) and UKIDSS $K$-band magnitude differences in Table~\ref{table.physical+optical}.
{The coordinates and individual magnitudes from where we computed the data are provided in Table~\ref{table.astro+photometry}.
In all but three cases,} the UKIDSS automatic pipeline was able to resolve both sources in at least the $K$ band.
Accounted UKIDSS {absolute} astrometric error varied between 0.05 and 0.07\,arcsec.
However, we considered the {relative} astrometric error at short angular separations to be lower, of the order of the photo-centroid uncertainty, which often is about one tenth of a pixel.
Similarly, accounted UKIDSS photometric errors of individual components varied between slightly less than one millimagnitude and two tenths of a magnitude, depending on signal-to-noise ratio, quality of PSF fit, and contamination by bright, close sources.

For the three exceptions for which {the UKIDSS automatic pipeline} was not able to resolve both components, namely Mayrit~359179\,AB, 785038\,AB and 1564349\,AB, $\rho$ and $\Delta K$ came instead from the separation of photo-centroids and ratio of flux peaks in the original {UKIDSS} images measured by us with the IRAF environment. 
Astrometric and photometric errors calculated with this method were 0.04--0.08\,arcsec and 0.03--0.04\,mag, respectively.
Besides, for the computation of $\Delta K$ of the bright binary \object{Mayrit~960106}, which saturates in the UKIDSS image, we used the 2MASS $K_{\rm s}$ magnitude of the primary as a reference.
We verified the automatic UKIDSS and our IRAF astrometric and photometric measurements with the IDL task {\tt xstarfinder}.
The new IDL measurements do not show any significant improvement with respect to the UKIDSS and IRAF ones, as previously warned by A.~Lawrence \& S.~Warren (priv.~comm.).

\begin{figure}
\centering
\includegraphics[width=0.49\textwidth]{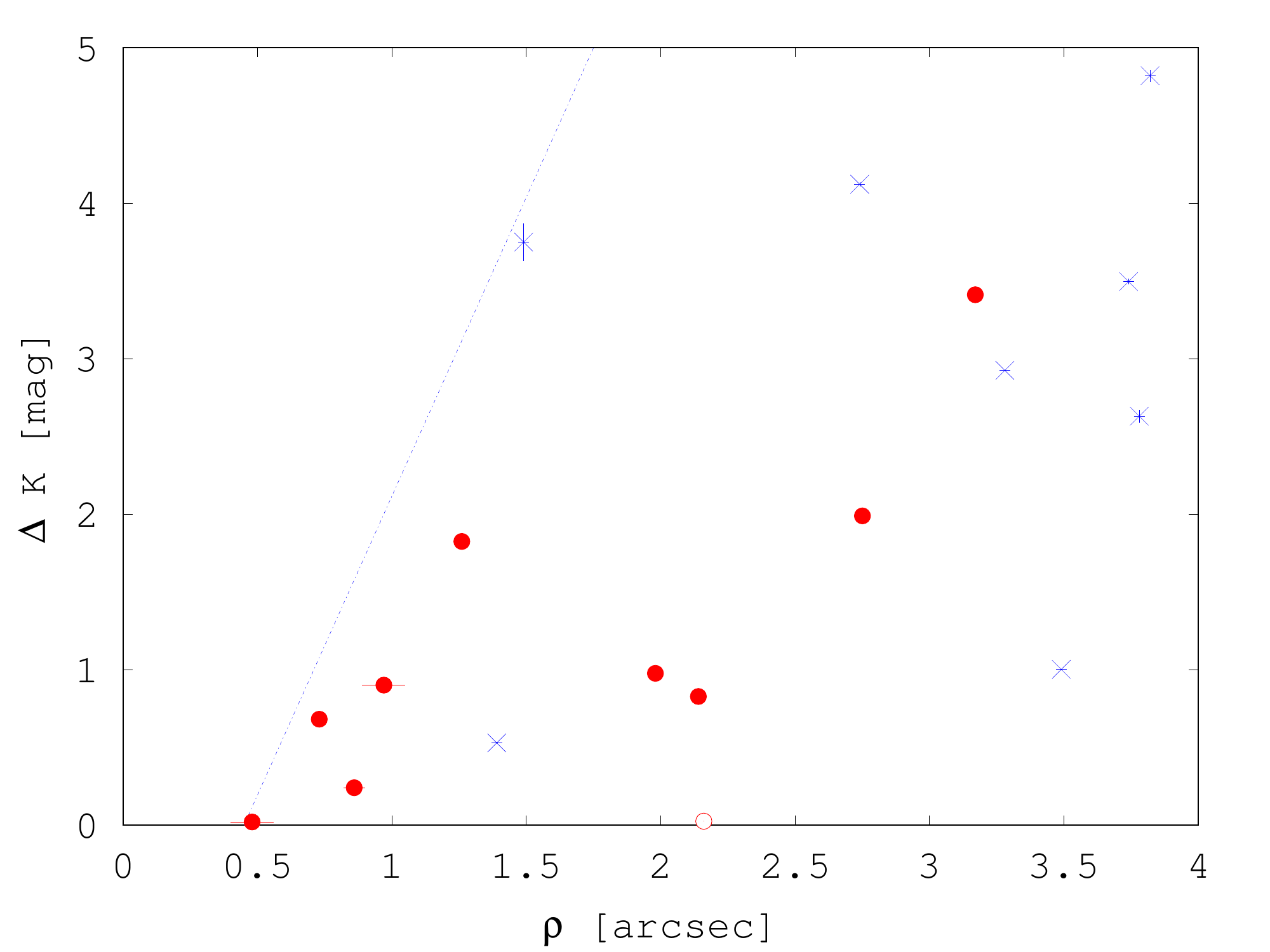}
\caption{$\Delta K$ vs. $\rho$ magnitude difference-angular separation diagram for the 18 pairs in Table~\ref{table.physical+optical}.
The dash-dotted line indicates the approximate detection limit of our survey.
Filled and open circles stand for physical pairs with primaries with and without known features of extreme youth, respectively, and crosses stand for optical {doubles} with secondaries in the cluster back- and foreground.
Errorbar sizes are of the order of symbol sizes.}
\label{figure.DKr}
\end{figure}

When investigating the companion detectability as a function of separation, $\Delta K$ vs. $\rho$, we concluded that it was easier to resolve the systems with {relatively large} angular separations, as expected, and with relatively faint components far from the saturation regime. 
In particular, we were able to resolve equal-brightness pairs (i.e. $\Delta K$ = 0\,mag) at angular separations $\rho \approx$ 0.5\,arcsec or larger, and to discover faint companions with magnitude differences $\Delta K$ = 5\,mag at $\rho \approx$ {1.7\,arcsec} or larger, {as illustrated in Fig.~\ref{figure.DKr}}.
 
Of the 18 identified multiple system candidates, eight had been discovered previously by the first author, but in different surveys (Caballero 2006, with WFC at the 2.5\,m Isaac Newton Telescope; Caballero 2007b, with CAIN-2 at the 1.5\,m Telescopio Carlos S\'anchez; Caballero 2008c, from asymmetries of stellar profiles in 2MASS images; Caballero 2010a and Caballero et~al. 2010b, with digitized photographic plates).
{Indeed, we have recovered all previously known `close' $\sigma$~Orionis binaries with angular separations in the interval 0.7\,arcsec $< \rho <$ 4.0\,arcsec, as tabulated by Caballero (2014).}
As shown in {his Table~III}, very few astrometric data had been provided for the eight {known} systems, but they approximately match with those measured by us here.

Most of the star and brown dwarf primaries in Table~\ref{table.physical+optical} have signposts of extreme youth (i.e., membership in $\sigma$~Orionis -- Caballero 2008c):
eight have strong Li~{\sc i} $\lambda$6707.8\,{\AA} lines in absorption in their spectra (among other youth features -- Muzerolle et~al. 2003; Caballero 2006; Sacco et~al. 2008; Hern\'andez et~al. 2014), one is a peculiar, magnetic, Herbig~Ae/Be star (Mayrit~960106: e.g., Bagnulo et~al. 2006), one is a strong H$\alpha$ emitter identified by four independent prism-objective surveys (\object{Mayrit~1245057}: Haro \& Moreno 1953; Wiramihardja et~al. 1989, 1991; Weaver \& Babcock 2004), one is a strong X-ray emitter discovered by {\em Einstein} (\object{Mayrit~1106058}: Wolk 1996; Caballero et~al. 2010b), and one has low surface gravity based on Na~{\sc i} $\lambda\lambda$8183.3,8194.8\,{\AA} absorption weaker than field dwarfs of the same spectral type and a circum(sub)stellar disc based on near-infrared flux excess (Mayrit~803197: Caballero et~al. 2007; Luhman et~al. 2008; Maxted et~al. 2008). 
Of the remaining six stars, one has a Li~{\sc i} equivalent width lower than expected for its magnitudes and colours (\object{Mayrit~785038}; but see Section~\ref{section.discussion}) and the other five are photometric cluster member candidates without spectroscopy (Scholz \& Eisl\"offel 2004; Sherry et~al. 2004; Caballero 2007a).
Lastly, L\'opez-Santiago \& Caballero (2008) associated a faint {\em Chandra} X-ray emission to {TYC~4770--1261--1}, which is unexpected for its estimated A--F spectral type; such X-ray emission may be caused by the ({optical}) companion candidate instead.

\begin{figure}
\centering
\includegraphics[width=0.49\textwidth]{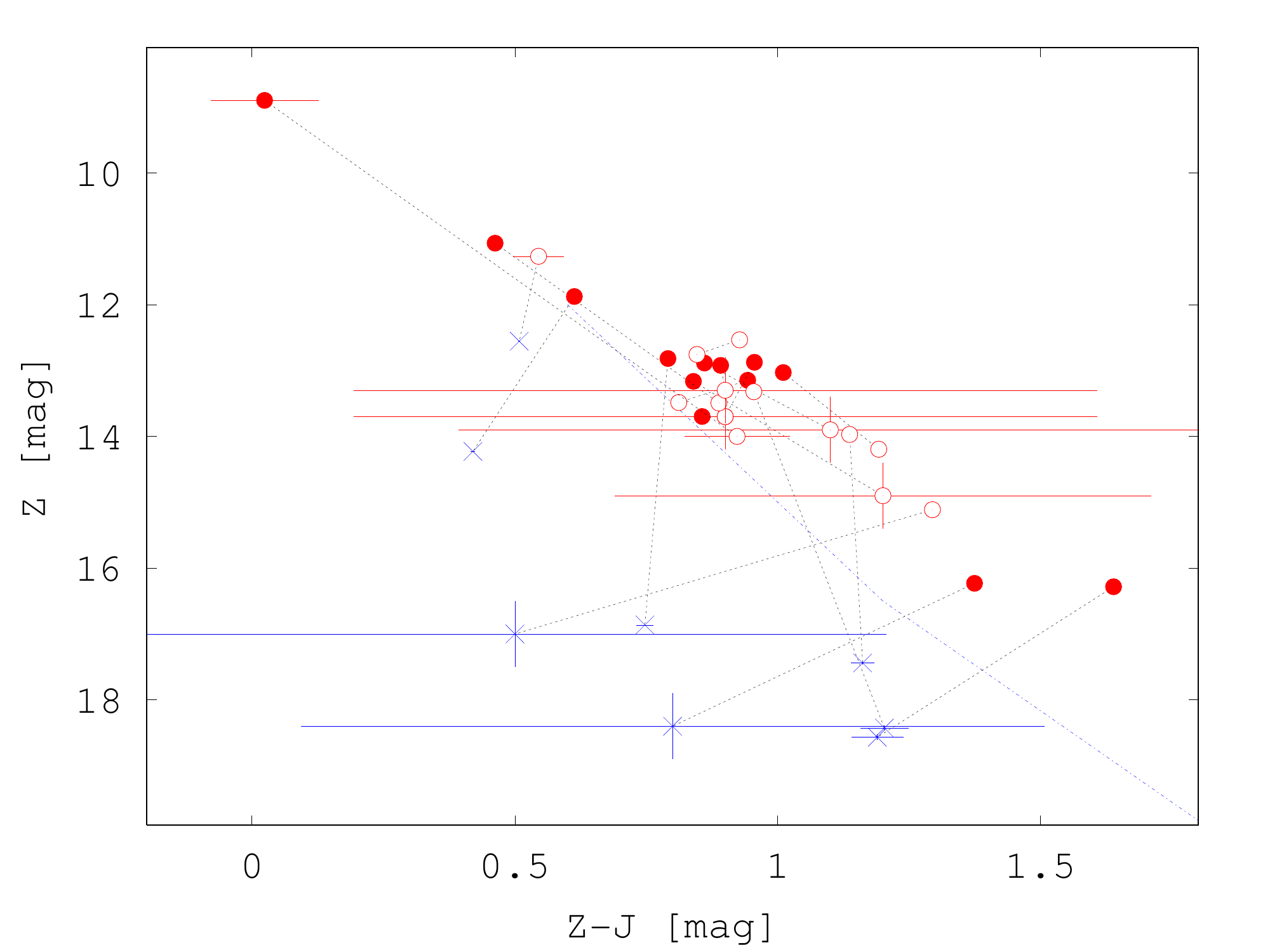}
\caption{$Z$ vs. $Z-J$ colour-magnitude diagram.
Same symbol meaning as in Fig.~\ref{figure.DKr}.
Objects redwards of the dash-dotted line are photometric cluster member candidates according to the Lodieu et~al. (2009) selection criterion in $Z$ and $J$ bands.
Some secondaries have large errorbars in $Z$.}
\label{figure.ZJ}
\end{figure}

{Ten} secondaries had accurate UKIDSS DR8+ photometry in at least the $ZYJK$ bands, which allowed us to apply the photometric selection criteria used by Lodieu et~al. (2009) in the $Z$\,vs.\,$Z-J$,  $Y$\,vs.\,$Y-J$, and $J$\,vs.\,$J-K$ colour-magnitude {diagrams}.
Of them, six had magnitudes inconsistent with membership in $\sigma$~Orionis.
We downloaded UKIDSS $Z$, $Y$, and $J$ images of the stars only resolved in $K$, and repeated the process of estimating $\Delta Z$, $\Delta Y$, and $\Delta J$ from ratios of flux peaks with IRAF {on the original UKIDSS images}.
In some cases, magnitude uncertainties were as high as 0.2--0.3\,mag, but small enough to discard another two very close ($\rho \sim$ 1.4--1.5\,arcsec) companion candidates to the young brown dwarf Mayrit~803197 and the photometric star cluster candidate Mayrit~1788137.
Fig.~\ref{figure.ZJ} illustrates this classification ($Y$ vs. $Y-J$ and $J$ vs. $J-K$ colour-magnitude diagrams do not provide extra information).

All the 20 primaries and secondaries in the upper part of Table~\ref{table.physical+optical} ({candidate physical binaries}) lie redwards of the Lodieu et~al. (2009) {photometric cluster member} selection criteria in $ZYJK$ bands {(see the $Z$ vs. $Z-J$ example in Fig.~\ref{figure.ZJ})}.
Of the ten primaries of candidate physical pairs, only one (\object{Mayrit~1411131}\,A) has no known feature of extreme youth. 
On the other hand, the eight systems with unbound companions, {which are listed in the bottom part of Table~\ref{table.physical+optical}, lie bluewards of of the Lodieu et~al.'s criteria and are, therefore, optical doubles in the classical nomenclature of double-star astronomers (e.g. Argyle 2012). 
In particular,} their secondaries are fainter than primaries, by over 2.5\,mag in all but two cases (which, if cluster members, would place them well in the substellar domain), but bluer or with the same approximate colours.
{Optical} companions are field late-type dwarf interlopers or, most likely, extragalactic sources in the cluster background (Caballero et~al. 2008a).

\section{Discussion}
\label{section.discussion}

\subsection{Physical nature of binary candidates}
\label{section.binarycandidates}

Since the proper motion of $\sigma$~Orionis is very low, of a few milliarcseconds per year (Orion is in the antapex and at about 400\,pc), currently we cannot discriminate between actual bound systems and chance alignments of pairs due to a projection effect without an appropriate radial-velocity study (Caballero 2010a).
However, Caballero (2009b) performed a statistical study and assigned to each possible pair in $\sigma$~Orionis a probability of random alignment as a function of the angular separations between the two components and between the primary and the cluster centre, which is defined by the massive {triple} OB system {$\sigma$~Ori\,Aa,Ab,B} (Caballero et~al. 2006 had done a less-elaborated statistical analysis before).
Such probabilities of alignment were calculated after simulating 1000 Monte Carlo spatial distributions of cluster members following the actual radial profile measured by Caballero (2008a) and computing angular separations between the nearest neighbours in each simulated cluster.
From Fig.~1 in Caballero (2009b), all of our physical binary candidates have {individual} probabilities of chance alignment less than 1\,\% (i.e., a probability of being physically bound greater than 99\,\%).
In the last column of Table~\ref{table.physical+optical}, we estimate new {individual} probabilities of chance alignment using the Caballero (2009b)'s Monte Carlo values and the Caballero (2008a)'s surface density of $\sigma$~Orionis members.
New values range from {0.5\,\%} for  \object{Mayrit~92149}\,AB, a pair with $\rho$ = 2.14$\pm$0.02\,arcsec and at 1.53\,arcmin to the cluster centre, to only 0.003\,\% for \object{Mayrit~785038}\,AB, a pair with $\rho$ = 0.48$\pm$0.08\,arcsec and at 13.1\,arcmin to the cluster centre.

\begin{figure}
\centering
\includegraphics[width=0.49\textwidth]{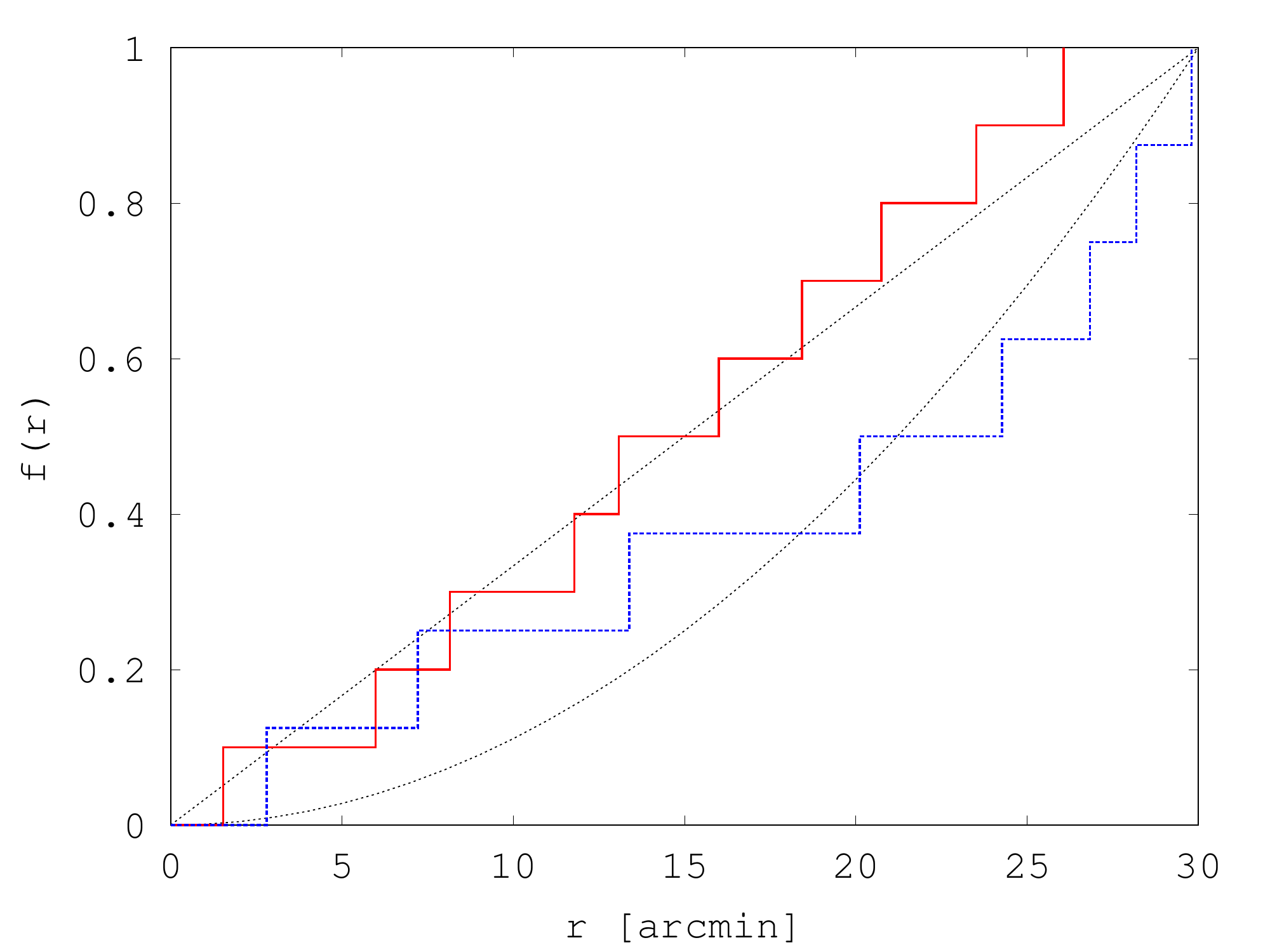}
\caption{Normalised cumulative number of systems with physical (red solid stair) and optical companions (blue dashed stair) as a function of the angular separation to the cluster centre, $f(r)$ vs. $r$, onto the power-law distributions for $f(r) \propto r^{1.0}$ and $r^{2.0}$ (black dotted lines), which correspond to the actual distributions of $\sigma$~Orionis members (top) and interlopers (bottom).
See Caballero (2008a) for details.}
\label{figure.fr}
\end{figure}

Moreover, although the statistics is low, the spatial distribution of physical binary candidates in Fig.~\ref{figure.fr} follows the power-law distribution of cluster members as in Caballero (2008a), while optical binary candidates follow the one of unrelated foreground/background sources, which are distributed uniformly within the 30\,arcmin-radius area centred on $\sigma$~Ori~Aa,Ab,B.
From now on, we will assume that {only} the ten pairs in the upper part of Table~\ref{table.physical+optical} actually orbit around a common centre of mass (i.e., are gravitationally bound), {while the other eight optical doubles are chance alignments.}

{A more accurate value of the contamination by chance alignments may come from calculating the change alignment probability for all the 331 primaries that we looked around in the 0.4--4.0\,arcsec separation interval and then sum them.
Such interval probabilities are strongly dependent on the separation to the cluster centre, with typical values of 10$^{-2}$, 10$^{-4}$, 10$^{-6}$\,\% at separations 1--10, 10--20, and 20--30\,arcsec, respectively.
After summing over the 331 primaries, we obtain a total probability of chance alignment of 1\,\%, approximately.
This total probability translates into expecting only about three interlopers in our survey, instead of the eight visual binaries found.
The reason for this apparent disagreement lies in the original estimation of individual probabilities of chance alignments by Caballero (2009), who took into account only random pairings of cluster stars, and not also of cluster stars with fore- and background sources.}

With the projected physical separations and individual masses estimated homogeneously from $J$ magnitudes and BT-Settl theoretical isochrones at 3\,Ma (Baraffe et~al. 2015), we computed reduced binding energies for the ten physical pairs as in Caballero (2009a). 
Values of $-U_g^*$ range in the interval from {130 to 2200\,10$^{33}$\,J} for \object{Mayrit~1564349}\,AB and Mayrit~960106\,AB, respectively.
These values are at least two orders of magnitude larger than the binding energies of the most fragile systems known in the field (at around 10$^{33}$\,J -- Weinberg et~al. 1989; Close et~al. 2003; Dhital et~al. 2010), which supports our assumption of physical binding.
{It is not surprising that we do not find physical binaries beyond the ``boundary between stellar kinematic groups and very wide binaries'' (Caballero 2009a) because we limited our pairing radius to only 4\,arcsec (see Section~\ref{section.analysis+results}).}

\subsection{Binary frequency}
\label{section.binaryfrequency}

At the cluster distance, the angular separations between {components in the ten physical pairs} translate into projected physical separations in the range $s$ = 180$\pm$30 (for Mayrit~785039\,AB) to 1220$\pm$80\,au (for Mayrit~960106\,AB), which are typically found in other open clusters at $d \le$ {400}\,pc (see Duch\^ene \& Kraus 2013 and references therein). 
The investigated interval of projected physical separations was of 160--1600\,au.

We measured a frequency of wide multiplicity in $\sigma$~Orionis of {3.0$^{+1.2}_{-1.1}$\,\% (10/331)}.
This frequency must be handle with care, since we were not sensitive to companions fainter by $\Delta K \gtrsim$ 5\,mag than their host stars or to any source at physical separations closer than $\sim$160\,au. 
Actually, from Fig.~\ref{figure.DKr}, we were not sensitive either to some companions between 160 and $\sim$650\,au ($\sim$1.7\,arcsec) depending on the magnitude difference with its primary.
Besides, we have not accounted for possible systems separated more than $\sim$1600\,au {because of the 4\,arcsec limit of our search.
In principle, wider companions fainter than the 2MASS completeness, but brighter than the UKIDSS completeness, should have been identified in previous deep photometric surveys (e.g. Lodieu et~al. 2009; Pe\~na-Ram\'{\i}rez et~al.~2012).}

Another possible effect to put the derived multiplicity rate into context is the Malmquist-like bias that leads to the preferential detection and identification of certain $\sigma$~Orionis members.
While the original Malmquist bias favours the detection of intrinsically bright objects, the identification of bona fide cluster members by all previous surveys in $\sigma$~Orionis has been biased towards young stars at all magnitudes with strong X-ray and/or H$\alpha$ emission, infrared-flux excess, or photometric variability (Caballero 2008c).
As it has been confirmed afterwards (see also Table~\ref{table.mayrit}), the degree of contamination in the Mayrit catalogue is very low, but there may be still some $\sigma$~Orionis members close to the substellar boundary yet to be discovered, especially with the bluest optical-to-infrared colours and lowest activity levels (cf. Caballero 2008c).
Thus, the derived overall multiplicity rate may be artificially high.

Regardless the effect of incompleteness in the 160--{650}\,au interval is larger or not than the Malmquist-like bias for the detection of active cluster members, a wide companion frequency $CF$ of {3.0$^{+1.2}_{-1.1}$\,\%} is quite low with respect to what has been published for other juvenile and young open clusters and star-forming regions (see the exhaustive reference list in Section~\ref{section.introduction}).
Unfortunately, given the heterogeneity in frequency derivations, cluster ages, surveyed primary masses, and physical/angular separations, it is difficult, if not impossible, to draw up any persuasive conclusion from the comparison.

As an example, the second lowest companion frequency after $\sigma$~Orionis' is that of the Orion Nebula Cluster ($CF$ = 8.8$\pm$1.1\,\% at $s$ = 68--680\,au; Reipurth et~al. 2007), which is slightly younger and has roughly the same heliocentric distance and metallicity (Gonz\'alez Hern\'andez et~al. 2008; Schaefer et~al. 2016).
{The apparently} lower $\sigma$~Orionis companion frequency can be ascribed to the lower masses of primaries (late-type, very low-mass stars have low binary frequencies; cf. Duch\^ene \& Kraus 2013 -- roughly 80--90\,\% of the Mayrit stars and brown dwarfs have K and M spectral types) and different surveyed projected physical separations (Padgett et~al. 1997, K\"ohler et~al. 2006 and Reipurth et~al. 2007 used the {\em Hubble Space Telescope}~or expensive adaptive optics systems, and were sensitive to closer companions).
{However, assuming a log-flat separation distribution, one gets a frequency of 6$\pm$2\,\% over the same separation range as Reipurth et~al., which is a little over 1$\sigma$ difference.}

Since the distribution of companions in open clusters peaks at about 3--5\,au (Patience et~al. 2002 -- see also the \"Opik's distribution), it is expected that a new high-resolution imaging survey for companions to the most massive $\sigma$~Orionis stars at angular separations shorter than 0.4\,arcsec ($s <$ 160\,au) would measure a companion frequency comparable to those of other open clusters.

\subsection{Miscellanea}
\label{section.miscellanea}

\subsubsection{Spectral types}
\label{section.SpT}

Only one star in the top of Table~\ref{table.physical+optical} has an early spectral type: the peculiar, magnetic, Herbig~Ae/Be star {Mayrit~960106}\,A (V1147~Ori), which is tabulated as an $\alpha^2$~CVn-type variable B9\,IIIp (Eu-Si) star (Joncas \& Borra 1981; North 1984; Catalano \& Renson 1998; Bagnulo et~al. 2006; Landstreet et~al. 2007).
Hern\'andez et~al. (2014) determined spectral types between M1.0 and M2.0 for another six primaries.
Except for Mayrit~1106058\,A, which is likely a K-type star, we estimate from {red optical and near-infrared} photometry that the other 12 stars (two primaries and ten secondaries) also lie in the M-type domain.

The latest spectral type corresponds to the faintest physical companion, \object{Mayrit~489165}\,B, which has an approximate $K$ magnitude of 13.3\,mag. 
This object, likely an M4--6 star, has a most probable mass just above the substellar boundary {(Caballero et~al. 2007)}.
Mayrit~489165\,B is also one of the three faintest companions with respect to their primaries.
The other two faint secondaries are Mayrit~1106058\,B ($\Delta K \sim$ 2.0\,mag) and Mayrit~960106\,B ($\Delta K \sim$ 3.4\,mag).
The remaining seven systems have differences $\Delta K <$ 1.0\,mag.
See Fig.~3 in Zapatero Osorio et al. (2003) for a magnitude vs. spectral type diagram of spectroscopically-confirmed members in $\sigma$~Orionis.

\subsubsection{A new Lindroos systems}
\label{section.Lindroos}

Large magnitude differences usually translate into small mass ratios ($q = \mathcal{M}_2/\mathcal{M}_1$).
Actually, the smallest mass ratio, of about {one tenth}, corresponds to the pair Mayrit~960106 (V1147\,Ori).
The peculiar B9 giant primary, with a mass of 3.5$^{+0.4}_{-0.3}$\,$M_\odot$ (Caballero 2007a), may form, together with its T~Tauri mid-M secondary, the second Lindroos system (Lindroos 1985) discovered in $\sigma$~Orionis after $\sigma$~Ori\,Aa,Ab,B and $\sigma$~Ori~IRS1\,AB (van~Loon \& Oliveira 2004; Caballero 2005; Hodapp et~al. 2009).
Furthermore, Caballero et~al. (2010b) remarked that {Mayrit~960106 and} Mayrit~524060 (an anonymous A8:\,V cluster star) were ``the only [early-type stars in $\sigma$~Orionis detected with HRC-I/{\em Chandra}] that are not known to form part of a multiple system''. 
We may wait for a dedicated study with {\em XMM-Newton} or, alternatively, a next generation of X-ray space missions (e.g., {\em Athena}), for ascertaining which is the actual origin of the X-ray emission in the Mayrit~960106 system: either the primary, previously thought to be single, or the {late-type} secondary at 3.17$\pm$0.02\,arcsec, which we propose here to be the true X-ray emitter.
The second scenario would explain the large $L_X/L_J$ ratio and angular separation between {\em Chandra} X-ray and 2MASS cross-matched sources found for Mayrit~960106 by Caballero et~al. (2010b).
(Today, {Mayrit}~524060 remains as the only {single} OBA-type cluster star with X-ray emission).

\subsubsection{X-rays and discs}
\label{section.X+discs}

Interestingly, eight of the ten physical pairs are known X-ray emitters, and the other two fall in areas not investigated in detail yet (Wolk 1996; Franciosini et~al. 2006; Skinner et~al. 2008; L\'opez-Santiago \& Caballero 2008).
Since roughly one half of the $\sigma$~Orionis stars have been detected with {\em Einstein}, {\em ROSAT}, {\em XMM-Newton}, and/or {\em Chandra} in X-rays (Caballero et~al. 2010b), it is hard to believe that this is a chance situation: the possibility of picking up eight X-ray stars among a sample of just eight random stars in the cluster (assuming an X-ray frequency of about 50\,\%) is $p \sim$ 0.4\,\%.
Therefore, the existence of companions at large projected physical separations {is able} to enhance the X-ray emission {in $\sigma$~Orionis}.
Components in young multiple systems cannot be tidally locked at $s \gtrsim$ 160\,au (tidal locking implies short rotation periods in close binaries, which translates into enhanced magnetic activity and coronal emission), but high-mass stellar wind collision, ionised mass trapped within long magnetic channels connecting the two low-mass stars, {unresolved spectroscopic binarity (thus making systems to be triple), or just a factor two in X-ray flux (no past or current X-ray space mission has resolved the systems in two components, both of which can be emitters)} can be other possible explanations.

On the other side, the minimum overall disc frequency in $\sigma$~Orionis is 30\,\% (Caballero et~al. 2007; Hern\'andez et~al. 2007; Luhman et~al. 2008).
The probability that there is no disc host system among our ten physical pairs, assuming such a disc frequency, is as low as 2.8\,\%. 
Thus, there is a {97.2\%} chance that the multiple systems have a lower disc frequency (but Mayrit~1245057 might have a disc;  Luhman et~al. 2008).
It could be even larger if one considers that each binary contains two stars, hence more chances of hosting at least one disc. 
{Although a larger sample is needed to assess its veracity, the existence of companions at large projected physical separations seems to affect or truncate circumstellar discs in $\sigma$~Orionis.}

\subsubsection{A remarkable pair}
\label{section.remarkable}

Mayrit~785038\,AB is, with $\rho$ = 0.48$\pm$0.08\,arcsec ($s$ = 180$\pm$30\,au), the closest pair in our sample.
It is an H$\alpha$ and X-ray emitter (Wiramihardja et~al. 1989; Franciosini et~al. 2006; Caballero et~al. 2010) and low-amplitude photometric variable (Scholz \& Eisl\"offel 2004), and has a radial velocity consistent with cluster membership (Sacco et~al. 2008).
It has, however, a lithium pseudo-equivalent width lower than expected for its spectral type, of $pEW$(Li~{\sc i}) = 133$\pm$7\,m{\AA} (with $\mathcal R \approx$ 17,000; Sacco et~al. 2007, 2008).
With spectral data of higher resolution ($\mathcal R \approx$ 34,000), Hern\'andez et~al. (2014) measured H$\alpha$ absorption and central emission and radial velocity compatible with previous values, but no lithium absorption. 
{However, typical $pEW$(Li~{\sc i})s of early and intermediate M dwarfs in $\sigma$~Orionis are 400--700\,m{\AA} (Zapatero Osorio et~al. 2002; Kenyon et~al. 2005; Caballero 2006; Gonz\'alez Hern\'andez et~al. 2008; Sacco et~al. 2008; Caballero et~al. 2012; Hern\'andez et~al. 2014).}

Far from attributing this severe lithium depletion to a post-T~Tauri phase, we ascribe it instead to the observed resolved multiplicity. 
For example, the first spectroscopic binary discovered in the $\sigma$~Orionis cluster, Mayrit~1415279\,AB (OriNTT~429), shows an apparent lithium depletion when observed at moderate resolution ($\mathcal R \approx$ 10,500; Caballero 2006), but the two lines are visible with \'echelle spectrographs at higher resolution (Lee et~al. 1994).
Assuming $M_A \approx M_B \approx$ 0.31\,$M_\odot$ {(again from $J$ magnitudes and NextGen models)} and $a \approx s$ (being $a$ the semi-major axis), the maximum radial-velocity semi-amplitude of one component in Mayrit~785038\,AB is $K_1 \sim 1.8 / \sqrt{1-e^2}$\,km\,s$^{-1}$.
A high eccentricity $e$ = 0.8 could lead to a separation of about 8\,km\,s$^{-1}$ in an epoch close to the periastron, which would easily blur the lithium line, as in the case of Mayrit~1415279\,AB.
However, Sacco et~al. (2007, 2008) did not report a significant veiling or line broadening in their spectra. 
Another possibility is that Mayrit~785038\,AB is actually an optical system of two mid-M dwarfs, one young (in $\sigma$~Orionis) and one old interloper (in the field), of very similar effective temperature, {apparent} brightness, and heliocentric radial velocity. 
Clearly, this pair deserves further detailed high-resolution spectroscopic and imaging studies.

\subsubsection{Go the limit}
\label{section.gothelimit}

During our survey, we also identified two stars with elongated PSFs in the northwest direction in the $K$ UKIDSS images.
We estimated angular separations $\rho \sim$ 0.2--0.3\,arcsec (projected physical separations $s \sim$ 80--120\,au at the $\sigma$~Orionis distance) and magnitude differences $\Delta K \lesssim$ 1\,mag in both cases, but were not able to estimate colours of the two secondaries.
The two very close pairs are Mayrit~1626148\,AB (V511~Ori, Haro 5--33) and Mayrit~873229\,AB (Haro 5--7).
The former is a photometric variable, young star with strong H$\alpha$ emission (Haro \& Moreno 1953; Fedorovich 1960; Wiramihardja et~al. 1991; Weaver \& Babcock 2004), while the latter has not only strong H$\alpha$ emission (Haro \& Moreno 1953; Hern\'andez et~al. 2014), but also near-infrared flux excess (Hern\'andez et~al. 2007), low surface gravity (Maxted et~al. 2008), and aperiodic photometric variability (Cody \& Hillenbrand 2010). 
Interestingly, Mayrit~873229 is also a double-line spectroscopic binary identified by Maxted et~al. (2008).
Because of the magnitude difference, we believe that the spectroscopic companion in Mayrit~873229 cannot be the same as the one identified in the UKIDSS image.
It is therefore a new candidate triple system (see Caballero 2014 for previously reported triples in $\sigma$~Orionis).

The other Mayrit stars, including {all known} spectroscopic binaries and candidates {compiled by Caballero (2014)}, have perfectly round PSFs in the UKIDSS images.
By including these two new {multiple} systems, if confirmed, the minimum frequency of wide multiplicity in the interval of projected physical separations $s$ = 80--1600\,au in $\sigma$~Orionis would be {3.6$^{+1.3}_{-1.2}$\,\%}.

\section{Conclusions}
\label{section.conclusions}

We searched for wide binaries in UKIDSS images of the nearby young $\sigma$~Orionis cluster.
We surveyed {331} cluster members and candidates in the corona from 1 to 30\,arcmin to the cluster centre and found 18 pairs with angular separations between 0.4 and 4.0\,arcsec.
We studied their membership in the cluster based on known youth features and relative $ZYJHK$ photometry.
Of them, eight are optical {doubles} which unbound secondaries are fore- or background sources.
The other ten pairs belong to the cluster, have {individual} probabilities of {chance} alignment $\le$1\,\%, follow the {photometric sequence and} spatial distribution of other cluster members, and, therefore, are very likely physical {(visual doubles)}.
Six of the ten bound binaries were reported by the first author in previous works, but had never been characterised homogeneously or in detail, while the other four are completely new.
The minimum frequency of wide multiplicity in the interval of projected physical separations $s$ = 160--1600\,au resulted in {3.0$^{+1.2}_{-1.1}$\,\%}.
{This companion frequency is low with respect to what has been measured in other open clusters and associations, such as the Pleiades, Upper Scorpius, or the Orion Nebula Cluster.
This fact may be due to the low mass of the primaries and that we have not been able to survey the closest regions of the stars, where the companion frequency increases.}
We found a trend of massive and X-ray stars to be in wide pairs or trios.
{Wide} multiplicity {seems} to affect the frequency of discs, {too}.
One of the new pairs, Mayrit~960106\,AB, is a Lindroos system of a peculiar Herbig Ae/Be star and a low-mass T~Tauri star, the latter of which my be the actual origin of the detected X-ray emission and flaring activity.
Our closest pair, Mayrit~785038\,AB ($\rho$ = 0.48$\pm$0.08\,arcsec), also new, had been reported to show a severe lithium depletion, which may be ascribed instead to unresolved binarity in spectra of moderate resolution.
Further spectroscopic studies with 8--10\,m-class telescopes, for confirming cluster membership based on irrefutable detection of lithium in absorption and common radial velocity, are needed to ascertain the true binding of our faint pairs.

Our programme using UKIDSS is not optimum for detecting faint companions at $\rho \lesssim$ 1\,arcsec to high- and mid-mass stars in $\sigma$~Orionis, but we paved the way for new, deeper, higher-resolution, imaging surveys.
This observational gap will be somewhat filled by an on-going lucky-imaging survey with FastCam at the 1.5\,m Telescopio Carlos S\'anchez (Oscoz et~al. 2008), which is unfortunately limited to the brightest stars in the cluster (R.~Rebolo, priv.~comm.). 

If we also account for the two extra multiple system candidates from UKIDSS data in Section~\ref{section.gothelimit} and the 15 known spectroscopic binaries, the five pairs wider than $\rho$ = 4\,arcsec outside the central arcminute and probability of alignment by chance $\le$1\,\%, and the two close binaries resolved with adaptive optics summarised by Caballero (2014), the {minimum} frequency of multiplicity in $\sigma$~Orionis rockets to about 10\,\%.
Since typical frequencies of multiplicity in young open clusters range between 30 and 40\,\% (cf. Duch\^ene \& Kraus 2013), it implies that $\sigma$~Orionis still harbours of the order of 80--100 unknown multiple systems (spectroscopic binaries and close multiples) that await discovery.

The ESA space mission {\em Gaia} will be able to discover tens of astrometric binaries in $\sigma$~Orionis, most of which will have periods of the order of the length of the mission.  
Unfortunately, neither the first data release has catalogued any close pair ({\em Gaia} Collaboration et~al. 2016), nor it is not expected that {\em Gaia} will {resolve} the three-dimensional structure of the cluster at almost 400\,pc, which may solve the alignment-by-chance problem of the widest pairs.
As a result, we encourage other authors to carry out {new} surveys for tight $\sigma$~Orionis binaries, both spectroscopic and with high-resolution imaging {at} large ground facilities (e.g., Very Large Telescope, Keck I and II, Gran Telescopio Canarias) or at the {\em Hubble} and {\em James Webb} spaces telescopes.


\acknowledgements
We thank {two anonymous referees for their reports}, V.~J.~S. B\'ejar for sharing unpublished results with us, H.~Bouy for helpful discussion on the $\sigma$~Ori system {and tools}, J.~Sanz-Forcada for advising about potential X-ray biases, and M.~R. Zapatero Osorio and R.~Rebolo for collecting adaptive optics data published by Caballero (2005). 
This research made use of the SIMBAD, operated at Centre de Donn\'ees astronomiques de Strasbourg, France, the NASA's Astrophysics Data System, {and the Washington Double Star Catalog}. 
Financial support was provided by the Spanish Ministerio de Econom\'{\i}a y Competitividad under grant AYA2014-54348-C3-2-R {and by the Klaus Tschira Stiftung}.


%


\appendix

\section{UKIDSS data and images}

   \begin{table*}
      \caption[]{UKIDSS coordinates and photometry of investigated pairs$^a$.} 
         \label{table.astro+photometry}
     $$ 
         \begin{tabular}{lc cc ccccc}
            \hline
            \hline
            \noalign{\smallskip}
Mayrit  	& 		& $\alpha$		& $\delta$			& $Z$				& $Y$				& $J$				& $H$				& $K$				\\
	  	& 		& (J2000)			& (J2000)			& [mag]				& [mag]				& [mag]				& [mag]				& [mag]				\\
            \noalign{\smallskip}
            \hline
            \noalign{\smallskip}
92149	& A		& 84.6995954		& --02.6220616	& 13.1480$\pm$0.0016	& 12.7705$\pm$0.0013	& 12.2054$\pm$0.0013	& 11.4491$\pm$0.0008	& 10.9082$\pm$0.0006	\\
		& B		& 84.7001288		& --02.6217988	& 13.4849$\pm$0.0019	& 13.1517$\pm$0.0016	& 12.6732$\pm$0.0016	& 11.9972$\pm$0.0011	& 11.7356$\pm$0.0010	\\
            \noalign{\smallskip}
359179	& A		& 84.6890597		& --02.6998671	& 12.8869$\pm$0.0014	& 12.5180$\pm$0.0012	& 12.0263$\pm$0.0012	& 11.5002$\pm$0.0008	& 11.0584$\pm$0.0007	\\
		& B		& ...    			& ...    			& ...    				& ...    				& ...    				&  ...$^a$				& ...    				\\
            \noalign{\smallskip}
489165	& A		& 84.7215567		& --02.7313454	& 13.1645$\pm$0.0016	& 12.8332$\pm$0.0013	& 12.3249$\pm$0.0013	& 11.8112$\pm$0.0010	& 11.5027$\pm$0.0009	\\
		& B		& 84.7215760		& --02.7309968	& ...    				& ...    				& ...    				& ...    				& 13.327$\pm$0.003	\\
            \noalign{\smallskip}
707162	& A		& 84.7463405		& --02.7870494	& 12.8756$\pm$0.0014	& 12.5067$\pm$0.0011	& 11.9201$\pm$0.0011	& 11.7037$\pm$0.0009	& 11.0587$\pm$0.0007	\\
		& B		& 84.7465423		& --02.7870470	& ...    				& ...    				& ...    				& ...    				& 11.7394$\pm$0.0010	\\
            \noalign{\smallskip}
785038	& A		& 84.8215539		& --02.4287051	& 13.6977$\pm$0.0019	& 13.4006$\pm$0.0017	& 12.8417$\pm$0.0017	& 12.1992$\pm$0.0013	& 11.9501$\pm$0.0012	\\
		& B		& ...    			& ...    			& ...    				& ...    				& ...    				& ...    				& ...    				\\
            \noalign{\smallskip}
960106	& A		& 84.9424723		& --02.6755447  	&  8.9$\pm$0.1$^b$		&  8.9$\pm$0.1$^b$		&  8.82$\pm$0.02$^b$	&  8.89$\pm$0.04$^b$	&  8.800$\pm$0.019$^b$	\\
		& B		& 84.9417748		& --02.6760830	& ...    				& 13.4054$\pm$0.0018	& 13.0774$\pm$0.0019	& ...    				& 12.2113$\pm$0.0013	\\
            \noalign{\smallskip}
1106058	& A		& 84.9475830		& --02.4378337	& 11.0673$\pm$0.0006	& 10.9849$\pm$0.0006	& 10.6050$\pm$0.0006	& 10.7019$\pm$0.0006	&  9.7292$\pm$0.0003	\\
		& B		& 84.9470918		& --02.4384202	& 13.4929$\pm$0.0017	& 13.1125$\pm$0.0015	& 12.6049$\pm$0.0015	& 11.9694$\pm$0.0011	& 11.7179$\pm$0.0010	\\
            \noalign{\smallskip}
1245057	& A		& 84.9761889		& --02.4105272	& 12.5345$\pm$0.0012	& 12.1524$\pm$0.0010	& 11.6072$\pm$0.0009	& 11.4639$\pm$0.0008	& 10.6698$\pm$0.0006	\\
		& B		& 84.9761443		& --02.4111269	& 12.7559$\pm$0.0013	& 12.5443$\pm$0.0011	& 11.9098$\pm$0.0011	& ...    				& 10.6947$\pm$0.0006	\\
            \noalign{\smallskip}
1564349	& A		& 84.6062412		& --02.1730235	& 12.9217$\pm$0.0014	& 12.57976$\pm$0.0012	& 12.0304$\pm$0.0012	& 11.4511$\pm$0.0009	& 11.1445$\pm$0.0007	\\
		& B		& ...    			& ...    			& ...    				& ...    				& ...    				& ...    				& ...    				\\
            \noalign{\smallskip}
            \hline
            \noalign{\smallskip}
168291	&  A 		& 84.6429755		& --02.5833693	& 11.8763$\pm$0.0009	& 11.7043$\pm$0.0008	& 11.2631$\pm$0.0008	& 11.0870$\pm$0.0007	& 10.4245$\pm$0.0005	\\
		& B		& 84.6437250		& --02.5828487	& 14.228$\pm$0.003	& 14.094$\pm$0.003	& 13.808$\pm$0.003	& 13.407$\pm$0.003	& 13.345$\pm$0.003	\\
            \noalign{\smallskip}
433123	& A		& 84.7872690		& --02.6661082	& 16.282$\pm$0.008	& 15.367$\pm$0.005	& 14.644$\pm$0.004	& 14.156$\pm$0.004	& 13.730$\pm$0.003	\\
		& B		& 84.7879204		& --02.6652836	& 18.57$\pm$0.04		& 18.05$\pm$0.03		& 17.38$\pm$0.03		& 16.62$\pm$0.03		& 16.36$\pm$0.04		\\
            \noalign{\smallskip}
803197	& A		& 84.6206707		& --02.8131332	& 16.231$\pm$0.008	& 15.506$\pm$0.005	& 14.857$\pm$0.005	& 14.333$\pm$0.005	& 13.908$\pm$0.004	\\
		& B		& 84.6202832		& --02.8132791	& ...    				& ...    				& ...    				& ...    				& 17.65$\pm$0.12	\\
            \noalign{\smallskip}
1207010	& A		& 84.7429939		& --02.2694735	& 12.8181$\pm$0.0013	& 12.5499$\pm$0.0012	& 12.0271$\pm$0.0011	& 11.6443$\pm$0.0009	& 11.0428$\pm$0.0007	\\
		& B		& 84.7436235		& --02.2690456	& 16.862$\pm$0.011	& 16.578$\pm$0.011	& 16.115$\pm$0.011	& 15.420$\pm$0.013	& 15.165$\pm$0.014	\\
            \noalign{\smallskip}
1456284	& A		& 84.2936993		& --02.5019597	& 11.2659$\pm$0.0007	& 11.3752$\pm$0.0007	& 10.9678$\pm$0.0007	& 11.0768$\pm$0.0007	& 10.4061$\pm$0.0005	\\ 
		& B		& 84.2945364		& --02.5014681 	& 12.5560$\pm$0.0012	& 12.3875$\pm$0.0011	& 12.0478$\pm$0.0012	& 11.6979$\pm$0.0009	& 11.4075$\pm$0.0008	\\
            \noalign{\smallskip}
1610344	& A		& 84.5605829		& --02.1709049	& 13.3222$\pm$0.0017	& 12.8971$\pm$0.0014	& 12.8334$\pm$0.0018	& 11.7905$\pm$0.0010	& 11.5054$\pm$0.0009	\\
		& B		& 84.5585370		& --02.1718970	& 19.17$\pm$0.07		& 18.98$\pm$0.08		& 18.83$\pm$0.12		& ...					& 18.3$\pm$0.2		\\
            \noalign{\smallskip}
1691180	& A		& 84.6824961		& --03.0698633	& 13.970$\pm$0.002	& 13.4245$\pm$0.0018	& 12.2054$\pm$0.0013	& 12.2098$\pm$0.0013	& 11.7906$\pm$0.0011	\\
		& B		& 84.6821895		& --03.0688691	& 17.440$\pm$0.019	& 16.942$\pm$0.013	& 16.278$\pm$0.012	& 15.625$\pm$0.016	& 15.286$\pm$0.017	\\
            \noalign{\smallskip}
1788137	& A		& 85.0282658		& --02.9607930	& 15.112$\pm$0.004	& 14.467$\pm$0.003	& 13.818$\pm$0.003	& 13.351$\pm$0.002	& 12.979$\pm$0.002	\\
		& B		& 85.0278811		& --02.9608262	& ...    				& ...    				& ...    				& 14.009$\pm$0.004	& 13.509$\pm$0.004	\\
        \noalign{\smallskip}
            \hline
         \end{tabular}
     $$ 
\begin{list}{}{}
\item[$^{a}$] {Ellipses for pairs not resolved by the automatic UKIDSS pipeline.}
\item[$^{b}$] {Mayrit~960106\,A $Z$ and $Y$ photometry affected by non-linearity and $J$, $H$ and $K$ (actually $K_s$) photometry taken from 2MASS.
Tabulated $Z$ and $Y$ photometry of Mayrit~1106058\,A and Mayrit~1456284\,A may also be affected by non-linearity.}
\end{list}
   \end{table*}

\begin{figure*}
\centering
\includegraphics[width=0.95\textwidth]{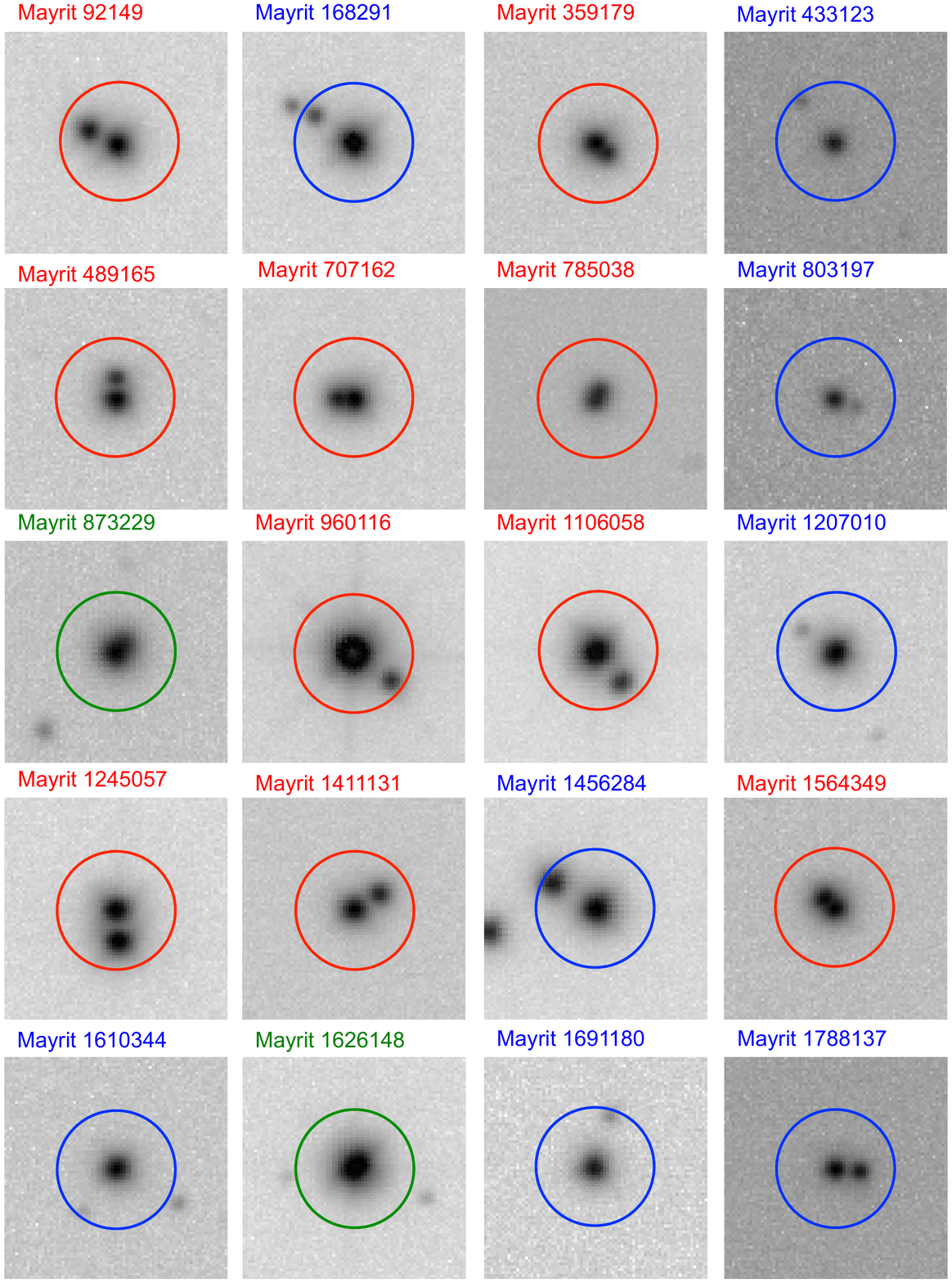}
\caption{Grey-inverted UKIDSS $K$-band images of the multiple systems investigated in this work (physical: red; optical: blue; with elongated PSFs: green).
Mayrit names are labeled on the top.
North is up, east is to the left.
Approximate size is 15\,arcsec.
Coloured circles centred on primaries are 4\,arcsec in radius.} %
\label{figure.estrellitas}
\end{figure*}

\end{document}